\documentclass[]{interact}

\usepackage{multirow}
\usepackage{hyperref}
\usepackage{algorithm}
\usepackage{algorithmic}
\usepackage{xcolor}
\usepackage{graphicx}
\usepackage{subfigure} 
\usepackage{epstopdf}
\usepackage{verbatim}

\usepackage[numbers,sort&compress]{natbib}
\bibpunct[, ]{[}{]}{,}{n}{,}{,}
\renewcommand\bibfont{\fontsize{10}{12}\selectfont}
\makeatletter
\def\NAT@def@citea{\def\@citea{\NAT@separator}}
\makeatother

\theoremstyle{plain}

\theoremstyle{definition}

\theoremstyle{remark}

\begin{document}


\title{Robust convex biclustering with a tuning-free method}

\author{
\name{Yifan Chen\textsuperscript{a}, 
Chunyin Lei \textsuperscript{b}, 
Chuanquan Li \textsuperscript{b,c*} \thanks{* CONTACT Chuanquan Li. Email: lichuanquan@jxufe.edu.cn}, 
Haiqiang Ma \textsuperscript{b,c},
Ningyuan Hu \textsuperscript{d}}
\affil{\textsuperscript{a}Department of Statistics and Applied Probability, University of California, Santa Barbara, United States; \\
\textsuperscript{b}School of Statistics and Data Science, Jiangxi University of Finance and Economics, Nanchang, China; \\
\textsuperscript{c}Key Laboratory of Data Science in Finance and Economics, Jiangxi University of Finance and Economics, Nanchang, China; \\
\textsuperscript{d}Department of Mathematics, University College London, Gower Street, London, WC1E 6BT, United Kingdom}
}

\maketitle

\begin{abstract}
Biclustering is widely used in different kinds of fields including gene information analysis, text mining, and recommendation system by effectively discovering the local correlation between samples and features. However, many biclustering algorithms will collapse when facing heavy-tailed data. In this paper, we propose a robust version of convex biclustering algorithm with Huber loss. Yet, the newly introduced robustification parameter brings an extra burden to selecting the optimal parameters. Therefore, we propose a tuning-free method for automatically selecting the optimal robustification parameter with high efficiency. The simulation study demonstrates the more fabulous performance of our proposed method than traditional biclustering methods when encountering heavy-tailed noise. A real-life biomedical application is also presented. The R package \texttt{RcvxBiclustr} is available at \href{https://github.com/YifanChen3/RcvxBiclustr}{https://github.com/YifanChen3/RcvxBiclustr}.
\end{abstract}

\begin{keywords}
Biclustering; Huber loss; Heavy tail; Tuning-Free; Convex optimization
\end{keywords}

\section{Introduction}

Biclustering, which was first proposed by Hartigan et al. \cite{hartigan1972direct}, tries to cluster rows and columns of a data matrix simultaneously. Lately, biclustering methods have been applied to a wide range of fields for data analysis and visualization such as biomedical data analysis, text mining, and recommendation system. To elaborate, in the domain of biomedical data analysis, researchers seek to identify patterns underlying the high-dimensional genetic data, which illustrate the local correlation between gene expression and patients, thus identifying subtypes of a certain disease \cite{lazzeroni2002plaid,bergmann2003iterative,turner2005biclustering,tan2014sparse,lee2010biclustering,chi2017convex}. In text mining, biclustering algorithms can recognize similar document subgroups by the local correlation between documents and words \cite{dhillon2001co,dhillon2003information}. In recommendation system, biclustering can be used to discover the local correlation between a group of certain customers and a particular category of products, and therefore improve the efficiency of recommendations \cite{hofmann1999latent,alqadah2015biclustering}. 

In this paper, we focus on proposing a biclustering algorithm that could be used to detect the local correlation in lung cancer microarray data regardless of the heavy-tailed noise. Lung cancer has several different subtypes. Our goal is to identify whether some genes have significant correlations with certain cancer subtypes. In other words, are some genes responsible for causing or restraining certain kinds of lung cancer? This is of great importance in both theoretical medical research and application. If we succeed in finding some significant correlations between genes and cancer subtypes, doctors can use this result as a reference to further determine personalized treatment strategies according to specific subtypes.

Recently, sparse biclustering has been increasingly popular in the field of statistics. Meanwhile, this kind of algorithm has three subgroups of derivatives: sparse biclustering based on K-means, sparse biclustering based on SVD, and convex biclustering. Tan and Witten \cite{tan2014sparse} extended the one-way K-means to two-way, thus proposing the SparseBC algorithm. However, since it is the derivative algorithm of one-way K-means, it shares the same limitation as one-way K-means. SparseBC takes a greedy approach, so it is sensitive to the initialization and can only reach a local optimal. Lee et al. \cite{lee2010biclustering} proposed a biclustering algorithm, SSVD, based on SVD innovatively. Sill et al. \cite{sill2011robust} adapted a robust tuning parameter based on SSVD and proposed the S4VD algorithm in consideration of data perturbance. Nevertheless, none of the methods above solves the two primary problems: sensitivity to initialization and local optimal. Chi et al. \cite{chi2017convex} proposed the convex biclustering algorithm (COBRA), which can be viewed as a convex optimization problem, therefore global optimal can be guaranteed and sensitivity to initialization exists no more. COBRA doesn’t need to designate the number of biclusters like SparseBC nor the number of layers like SSVD. It only has one tuning parameter, the coefficient of penalty term, $\lambda$. Consequently, the process of tuning the parameter is relatively trivial and efficient. 

However, the effectiveness of COBRA will significantly reduce when the data contains heavy-tailed noise. Specifically, the distribution of a random variable $X$ with distribution function $F$ is said to have a heavy tail if the moment generating function of $X$, $M_X(t)$, is infinite for all $t > 0$, namely, $\int_{-\infty}^{+\infty}e^{tx}dF(x)=\infty$ for all $t>0$. Moreover, heavy-tailed noise is inevitable in high-dimensional data, so it is unrealistic to expect high-dimensional data in the fields of finance, macroeconomics, and biomedical to have sub-Gaussian distribution \cite{cont2007volatility,stock2002forecasting,pan2017resveratrol,balcan2014robust,suzek2015uniref}. Huber loss \cite{huber1992robust} was proposed to mitigate the negative influence of outliers with the following form:

\begin{equation}
    \mathcal{L}_\tau(a)=\left\{
    \begin{array}{rcl}
         &\frac{1}{2}a^2, &|a|\leq \tau\\
         &\tau|a|-\frac{1}{2}\tau^2, &|a|>\tau
    \end{array}
    \right.
\end{equation} Notice that Huber loss is exactly squared error loss when $|a|\leq\tau$. The robustness comes from the part that $|a|>\tau$, where the absolute error loss serves as a buffer that controls the rising speed of squared error loss. Huber regression is also proposed to address the heavy-tailed noise problem \cite{huber1973robust}. Yohai and Maronna \cite{yohai1979asymptotic}; Mammen \cite{mammen1989asymptotics}; He and Shao \cite{he2000parameters} did thorough research on the asymptotic properties of M-estimators. On the other hand, for the non-asymptotic point of view, we refer to \cite{catoni2012challenging,tan2022sparse,ke2019user,sun2020adaptive}. Liu et al. \cite{liu2019robust} proposed robust convex clustering, which adapted the cost function and weight function of convex clustering into Huber loss, thus making it perform well when facing data containing heavy-tailed noise. Based on this thought, we extend this idea into convex biclustering by replacing the squared error term in the cost function and weight function with Huber loss. 

Yet, this adaptation introduces a new parameter, the robustification parameter in Huber loss, $\tau$, making not only the process of tuning the parameters time-consuming but also the parameter hard to determine. Sun et al. \cite{sun2020adaptive} proposed the adaptive Huber regression for robust estimation and inference whose robustification parameter is adapted to the sample size, dimension, and moments. Wang et al. \cite{wang2021new} proposed a tuning-free method for Huber regression. Accordingly, we extend this idea into our method and put the tuning of $\tau$ into the iteration of the algorithm. Therefore, the proposed algorithm won’t be needing to tune the parameter $\tau$ manually when it comes to simulation and application, which speeds up the process and increases the accuracy.

The rest of this paper is organized as follows. In Section~\ref{sec-algo}, we propose a Robust Convex BiClustering algorithm (RCBC). Section~\ref{sec-tp} proposes a tuning-free method for selecting a desirable $\tau$ for the algorithm, and demonstrates the cross-validation procedure we use to select the penalty parameter $\lambda$. Section~\ref{sec-sim} contains the result of our simulation study. In Section~\ref{sec-app}, we implement RCBC on gene data for an application. And finally, the discussion is in Section~\ref{sec-dis}.

\section{Robust Convex Biclustering}
\label{sec-algo}

We seek to identify a checkboard pattern that reflects the local correlation between the features and the samples. Let $\mathbf{X} \in \mathbb{R}^{n\times p}$ be a data matrix with $n$ samples and $p$ features and $\mathbf{U} \in \mathbb{R}^{n\times p}$ be the estimated matrix generated by minimizing the following convex optimization problem

\begin{equation}
    F_{\lambda,\tau}(\mathbf{U})=\mathbf{L}_\tau(\mathbf{X}-\mathbf{U})+\lambda\left[\Omega_\mathbf{W}(\mathbf{U})+\Omega_{\Tilde{\mathbf{W}}}(\mathbf{U}^T)\right]
\end{equation} where $\mathbf{L}_\tau(\mathbf{X}-\mathbf{U})=\sum\limits_{i=1}^n\sum\limits_{j=1}^p \mathcal{L}_\tau(\mathbf{X}_{ij}-\mathbf{U}_{ij})$, $\Omega_\mathbf{W}(\mathbf{U})=\sum_{i<j}w_{ij}||\mathbf{U}_{\cdot i}-\mathbf{U}_{\cdot j}||_2$, and $\mathbf{U}_{\cdot i} (\mathbf{U}_{i\cdot})$ denotes the $i$th column(row) of the matrix $\mathbf{U}$. The above equation is a fused lasso problem in which the cost function represents how well $\mathbf{U}$ approximates $\mathbf{X}$ and the penalty term tends to make close elements fuse with each other from both directions. The nonnegative tuning parameter $\lambda$ controls the tradeoff between the cost function and the penalty term. When $\lambda=0$, $F_{\lambda,\tau}(\mathbf{U})$ reaches its optimal when $\mathbf{U}=\mathbf{X}$. As $\lambda$ increases, the close elements begin to coalesce, and the checkboard pattern starts to merge. When $\lambda$ is sufficiently large, $\mathbf{U}$ will tend to shrink into one bicluster with all the entries being a constant. 

The cost function is a robust adaptation of squared error loss. In real life, data cannot be perfect, and heavy-tailed noise, which has a significant negative influence on squared error loss, is inevitable. Therefore, a robust version of convex biclustering will provide us with wider application fields compared to the normal version. We choose the Huber loss \cite{huber1992robust} to address the heavy-tailed noise.

The penalty term is the same as convex biclustering \cite{chi2017convex}, and the weight function is also adapted with Huber loss as follows:

\begin{equation}
    w_{ii^*}^k=I_{(i,i^*)}^k \mathcal{\exp}\left\{-\xi\left[\sum\limits_{j\in D_1}(\mathbf{X}_{i^*j}-\mathbf{X}_{ij})^2+\sum\limits_{j\in D_2}\delta^2\right]\right\}
\end{equation} where $D_1=\{j:|\mathbf{X}_{i^*j}-\mathbf{X}_{ij}|\leq\delta\}$, $D_2=\{j:|\mathbf{X}_{i^*j}-\mathbf{X}_{ij}|>\delta\}$, and $I_{(i,i^*)}^k$ is the indicator term which is 1 if $i^*$ is one of $i$'s $k$ nearest neighbours and 0 otherwise. Note that by implementing the indicator function, we successfully make the weight function sparse. The indicator function $I_{(i,i^*)}^k$ here significantly increases the efficiency and accuracy of the algorithm by ignoring most of the useless uncorrelated rows(columns) and making the weight function sparse. Likewise, simply substitute $\mathbf{X}$ with $\mathbf{X}^T$ can we get $\Tilde{w}_{ii^*}^k$. This weight function has a valuable property that assigns greater weights to contaminated rows(columns) that belong to the same cluster than to rows(columns) that belong to different clusters. In the later simulation and application sections, we set $\xi=0.001$ and $\delta=1.345\hat{\sigma}$, where $\hat{\sigma}$ denotes the median absolute deviation (MAD) estimator of $\mathbf{X}$, as default.

After analysing the formula of robust convex biclustering, we now come to the phase of solving it. Notice that if the penalty term only has regularization in one direction instead of penalizing both rows and columns, the solution will be much easier with the Alternating Direction Method of Multipliers (ADMM) \cite{boyd2011distributed}. However, penalizing both directions is of necessity. Dykstra-like proximal algorithm (DLPA) \cite{bauschke2008dykstra} was proposed to tackle this kind of problem by iteratively solving two convex optimization problems.

Typically, DLPA solves problems with the form

\begin{equation}
    \mathop{\min}_{\mathbf{U}} \frac{1}{2}||\mathbf{X}-\mathbf{U}||_F^2+f(\mathbf{U})+g(\mathbf{U})
\end{equation} where $f$ and $g$ are lower semicontinuous and convex function, and $||\cdot||_F$ is the Frobenius norm. Let $f=\lambda\Omega_\mathbf{W}$, $g=\lambda\Omega_{\Tilde{\mathbf{W}}}$. The extension from squared error loss to Huber loss is natural as the latter is also convex. The proximal mapping in the original DLPA algorithm thus becomes

\begin{equation}
    \mathop{\arg\min}_{\mathbf{U}} \mathbf{L}_\tau(\mathbf{X}-\mathbf{U})+\lambda\Omega_\mathbf{W}(\mathbf{U})
\end{equation} which happens to be the formula of robust convex clustering.

Therefore, the thought of our algorithm is straightforward. Informally, we just perform one-way robust convex clustering algorithms to rows and columns respectively in one iteration and then iterate until convergence. Yet, some details still need to be clarified. Let $\mathbf{U}^{(m)}$ and $\mathbf{R}^{(m)}$ be the estimated mean matrix at the $m$th iteration of $\mathbf{X}$ and $\mathbf{X}^T$ respectively. $\mathbf{P}^{(m)}$ reflects the discrepancy between $\mathbf{U}^{(m)}$ and $\mathbf{R}^{T(m)}$, and likewise, $\mathbf{Q}^{(m)}$ reflects the discrepancy between $\mathbf{U}^{T(m)}$ and $\mathbf{R}^{(m)}$. To make things clear, $\mathbf{U}^{(m)}$, $\mathbf{P}^{(m)} \in \mathbb{R}^{n\times p}$ while $\mathbf{R}^{(m)}$, $\mathbf{Q}^{(m)} \in \mathbb{R}^{p\times n}$. Instead of simply clustering rows and columns in each iteration, we actually cluster $\mathbf{U}^{(m)}+\mathbf{P}^{(m)}$ and $\mathbf{R}^{(m)}+\mathbf{Q}^{(m)}$. Additionally, the stop criterion is trivial. The iteration will stop until $||\mathbf{U}^{(m)}-\mathbf{R}^{T(m)}||_F < \varepsilon$, where $\varepsilon >0$ is the tolerance of this iteration which is efficiently small. The pseudocode of our algorithm is shown in Algorithm~\ref{alg-rcbc}.

\begin{algorithm}
	\renewcommand{\algorithmicrequire}{\textbf{Input:}}
	\renewcommand{\algorithmicensure}{\textbf{Output:}}
	\caption{Robust Convex Biclustering Algorithm}
	\label{alg-rcbc}
	\begin{algorithmic}[1]
		\STATE Initialization: $\mathbf{U}^{(0)}\leftarrow\mathbf{X},\mathbf{R}^{(0)}\leftarrow\mathbf{X}^T,\mathbf{P}^{(0)}\leftarrow\mathbf{0},\mathbf{Q}^{(0)}\leftarrow\mathbf{0},m\leftarrow 0$
		\REPEAT
		\STATE $m \leftarrow m + 1$
		\STATE $\mathbf{R}^{(m)} \leftarrow \texttt{Robust\_Convex\_Clustering} (\mathbf{U}^{T(m-1)}+\mathbf{P}^{T(m-1)})$
		\STATE $\mathbf{P}^{(m)} \leftarrow \mathbf{P}^{(m-1)}+\mathbf{U}^{(m-1)}-\mathbf{R}^{T(m)}$
		\STATE $\mathbf{U}^{(m)} \leftarrow \texttt{Robust\_Convex\_Clustering} (\mathbf{R}^{T(m)}+\mathbf{Q}^{T(m-1)})$
  		\STATE $\mathbf{Q}^{(m)} \leftarrow \mathbf{Q}^{(m-1)}+\mathbf{R}^{(m)}-\mathbf{U}^{T(m)}$
		\UNTIL $||\mathbf{U}^{(m)}-\mathbf{R}^{T(m)}||_F  < \varepsilon $
		\ENSURE  $\mathbf{U}^{(m)}$
	\end{algorithmic}  
\end{algorithm}

The one-way robust convex clustering algorithm is almost the same as Liu et al. \cite{liu2019robust} proposed except for the weight function which is adjusted to be sparse as we mentioned before. Typically, this one-way algorithm solves the constrained optimization problem

\begin{equation}
    \begin{aligned}
        & \mathop{\min}_{\mathbf{U},\mathbf{W}\in \mathbb{R}^{n\times p},\mathbf{V} \in \mathbb{R}^{{n \choose 2}\times p}} \mathbf{L}_\tau(\mathbf{X}-\mathbf{W})+\lambda\sum_{i<j}w_{ij}||\mathbf{V}_{ij}||_2 \\
        &  \text{subject to}\ \ \mathbf{U}_i=\mathbf{W}_i \ \  \text{and}\ \ \mathbf{U}_i-\mathbf{U}_j=\mathbf{V}_{ij} \ \  \text{for all}\ \ i<j
    \end{aligned}
\end{equation} where $\mathbf{V}$ is an ${n\choose 2}\times p$ matrix, and $\mathbf{V}_{ij}$ denotes the $ij$th row of matrix $\mathbf{V}$. Let $\mathbf{E}$ be an ${n\choose 2}\times n$ matrix and $(\mathbf{EU})_{ij}=\mathbf{U}_i-\mathbf{U}_j$. Consequently, the corresponding augmented Lagrangian function will be

\begin{equation}
    \begin{aligned}
        F(\mathbf{U},\mathbf{W},\mathbf{V},\mathbf{Y},\mathbf{Z})&=\mathbf{L}_\tau(\mathbf{X}-\mathbf{W})+\lambda\sum_{i<j}w_{ij}||\mathbf{V}_{ij}||_2\\
        &+\frac{\rho}{2}||\mathbf{V-EU+Y}||_F^2+\frac{\rho}{2}||\mathbf{W-U+Z}||_F^2
    \end{aligned}
    \label{formula-lag}
\end{equation} where $\mathbf{U,W,V}$ are primal variables, $\mathbf{Y}\in \mathbb{R}^{{n\choose 2} \times p},\mathbf{Z} \in \mathbb{R}^{n\times p}$ are dual variables, and $\rho$ is a nonnegative tuning parameter for the ADMM algorithm. The update procedure for $\mathbf{U},\mathbf{W},\mathbf{V},\mathbf{Y}$, and $\mathbf{Z}$ is as follows:

\begin{equation}
    \begin{aligned}  
        \mathbf{U}^{(m)}&=\left(\mathbf{E}^T\mathbf{E}+\mathbf{I}\right)^{-1}\left[\mathbf{E}^T\left(\mathbf{V}^{(m-1)}-\mathbf{Y}^{(m-1)}\right)+\mathbf{W}^{(m-1)}+\mathbf{Z}^{(m-1)}\right] \\
        W_{ij}^{(m)}&=\left\{
            \begin{array}{rcl}
            &\left[X_{ij}+\rho\left(U_{ij}^{(m)}-Z_{ij}^{(m-1)}\right)\right]/(1+\rho), &\frac{\rho}{1+\rho}\left|X_{ij}-\left(U_{ij}^{(m)}-Z_{ij}^{(m-1)}\right)\right|\leq \tau\\
            &X_{ij}+soft\left(U_{ij}^{(m)}-Z_{ij}^{(m-1)}-X_{ij},\frac{\tau}{\rho}\right), &o.w.
            \end{array}
        \right. \\
        \mathbf{V}_{ij}^{(m)}&=\left[1-\frac{\lambda w_{ij}}{\rho ||\mathbf{U}_i^{(m)}-\mathbf{U}_j^{(m)}-\mathbf{Y}_{ij}^{(m-1)}||_2}\right]_{+}\left(\mathbf{U}_i^{(m)}-\mathbf{U}_j^{(m)}-\mathbf{Y}_{ij}^{(m-1)}\right) \\
        \mathbf{Y}_{ij}^{(m)}&=\mathbf{Y}_{ij}^{(m-1)}-\rho \left(\mathbf{U}_i^{(m)}-\mathbf{U}_j^{(m)}-\mathbf{V}_{ij}^{(m)}\right) \\
        \mathbf{Z}^{(m)}&=\mathbf{Z}^{(m-1)}-\rho \left(\mathbf{U}^{(m)}-\mathbf{W}^{(m)}\right)
    \end{aligned}
    \label{formula-one}
\end{equation} where $soft(a,b)=sign(a)\mathcal{\max}(|a|-b,0)$ is the the soft-thresholding operator, and $[a]_{+}=\mathcal{\max}(a,0)$. The detailed derivation of the one-way algorithm can be found in Appendix~\ref{appendix}.

\section{Tuning Parameters}
\label{sec-tp}
\subsection{Tuning-Free Method for Selecting $\tau$}
\label{subsec-tf}

The proposed RCBC has a convincing performance in dealing with heavy-tailed noise. However, the newly introduced robustification parameter $\tau$ brings challenges to tuning the hyperparameters with two-dimensional grid search and cross-validation, as tuning two parameters will be much more computationally demanding compared to just tuning one. In order to speed up the process of selecting the optimal parameters, we proposed a tuning-free method for selecting $\tau$.

An empirical selection for $\tau$ in Huber's original proposal is $\tau=1.345\sigma$, which retains 95\% of the asymptotic efficiency of the estimator for normally distributed data \cite{huber2011robust}. Our proposed $\tau$ is more correlated to the data thus being called data-driven and adaptive. In robust regression problems, a desirable $\tau$ should adapt to sample size $n$, dimension $p$, and moments $v$ for an optimal trade-off between bias and robustness \cite{sun2020adaptive}. Our biclustering problem can also be viewed as a high-dimensional regression problem solved by fused lasso.

\begin{equation}
    \mathbf{L}_\tau\left(vec(\mathbf{X})-\mathbf{D}\times vec(\mathbf{U})\right)+\lambda\left[\sum_{i<j}w_{ij}||\mathbf{U}_{\cdot i}-\mathbf{U}_{\cdot j}||_2+\sum_{i<j}\Tilde{w}_{ij}||\mathbf{U}^T_{\cdot i}-\mathbf{U}^T_{\cdot j}||_2\right]
\end{equation} where $vec(\mathbf{X}), vec(\mathbf{U}) \in \mathbb{R}^{np\times 1}$ are vectors that come from stretching the matrices, and $\mathbf{D} \in \mathbb{R}^{np\times np}$ is an identity matrix that serves as the design matrix of the regression problem.

Therefore, the formation of our biclustering problem is almost the same as the high-dimensional adaptive Huber regression in \cite{wang2021new} which is guided by non-asymptotic deviation analysis. Accordingly, we add a procedure for automatically updating $\tau$ in the original RCBC algorithm, namely updating $\tau$ and solving one-way clustering problems of two directions simultaneously in one iteration. Using the previous estimated $\mathbf{U}^{(m-1)}$, we compute $\tau^{(m)}$ as the solution to

\begin{equation}
    \frac{1}{np - s^{(m-1)}}\sum\limits_{i=1}^{np}\frac{\mathop{\min}\{(vec(\mathbf{X})_i-vec(\mathbf{U}^{(m-1)})_i)^2,\tau^2\}}{\tau^2}=\frac{\log(np\times np)}{np}
    \label{equ-tuning-free}
\end{equation} where $s^{(m-1)}=\mathop{\min}\{||\mathbf{V}||_{0^r},||\Tilde{\mathbf{V}}||_{0^r}\}, \mathbf{V} \in \mathbb{R}^{{n\choose 2} \times p}, \Tilde{\mathbf{V}} \in \mathbb{R}^{{p\choose 2} \times n}$, and $||\cdot||_{0^r}$ represents the number of rows of a matrix that not all elements are 0.  See the detailed derivation and desirable statistical properties of Equation~\ref{equ-tuning-free} in \cite{wang2021new}.

The pseudocode for this tuning-free version of RCBC is given in Algorithm~\ref{alg-tf}.

\begin{algorithm}
	\renewcommand{\algorithmicrequire}{\textbf{Input:}}
	\renewcommand{\algorithmicensure}{\textbf{Output:}}
	\caption{Robust Convex Biclustering with Tuning-Free Method}
	\label{alg-tf}
    \begin{algorithmic}[1]
		\STATE Initialization: $\mathbf{U}^{(0)}\leftarrow\mathbf{X},\mathbf{R}^{(0)}\leftarrow\mathbf{X}^T,\mathbf{P}^{(0)}\leftarrow\mathbf{0},\mathbf{Q}^{(0)}\leftarrow\mathbf{0},m\leftarrow 0$
		\REPEAT
		\STATE $m \leftarrow m + 1$
		\STATE Update $\mathbf{R}^{(m)},\mathbf{P}^{(m)},\mathbf{U}^{(m)},\mathbf{Q}^{(m)}$ as Algorithm 1
        \STATE Update $\tau^{(m)}$ based on Equation (9)
    
		\UNTIL $||\mathbf{U}^{(m)}-\mathbf{R}^{T(m)}||_F  < \varepsilon $
		\ENSURE  $\mathbf{U}^{(m)}$
	\end{algorithmic}
\end{algorithm}

\subsection{Tuning $\lambda$ with Missing Data}
\label{subsec-cv}

Our proposed RCBC algorithm has only one tuning parameter $\lambda$, which is also our advantage in the process of selecting the tuning parameter compared to other algorithms. As for the method of selecting the parameter of this unsupervised problem, we refer to the classic missing data method \cite{witten2009penalized,tan2014sparse,chi2017convex,xu2021integrative}, which can recast biclustering as a supervised learning problem. Here, it is detailed as Algorithm~\ref{alg-t-para}:

\begin{algorithm}
	\renewcommand{\algorithmicrequire}{\textbf{Input:}}
	\renewcommand{\algorithmicensure}{\textbf{Output:}}
	\caption{Selecting tuning parameter $\lambda$ with missing data}
	\label{alg-t-para}
    \begin{algorithmic}[1]
        \FORALL{candidate $\lambda$}
            \FOR{k = 1 : T}
                \STATE Randomly set $np/T$ elements of the matrix to be missing
                \STATE Impute the mean of non-missing values to construct a new matrix $\mathbf{X}^k$
                \STATE Perform our proposed algorithm on $\mathbf{X}^k$
                \STATE Calculate the mean squared error
                $$\sum\limits_{\mathbf{X}_{ij}\ \text{is}\ \text{missing}}\frac{(\mathbf{X}_{ij}-\mathbf{M}_{ij})^2}{np/T}$$
                where $\mathbf{M}$ is the estimated bicluster mean matrix
            \ENDFOR
    		\STATE Calculate the average of the $T$ mean squared errors $MSE_\lambda$
        \ENDFOR
        \STATE Select the optimal $\lambda$ based on the lowest $MSE_\lambda$
	\end{algorithmic}
\end{algorithm}

Normally, $T$ in Algorithm~\ref{alg-t-para} is set to be 10 to perform 10-fold cross-validation.

\section{Simulation}
\label{sec-sim}

In this section, we compare the performance of our tuning-free biclustering algorithm (TF-RCBC), our non-tuning-free biclustering method (RCBC), COBRA \cite{chi2017convex}, and spBC \cite{tan2014sparse} in ten simulation settings. Our proposed methods are available as an R package at \href{https://github.com/YifanChen3/RcvxBiclustr}{https://github.com/YifanChen3/RcvxBiclustr}. COBRA and spBC can be implemented in R packages \textbf{\texttt{cvxbiclustr}} and \textbf{\texttt{sparseBC}} respectively on CRAN.

To evaluate the quality of the clustering results, we use the following metrics: the Rand index (RI), the adjusted Rand index (ARI), and the variation of information (VI). The RI \cite{rand1971objective} is the most popular measure of the similarity between two data clusterings. Nevertheless, the RI can't ensure a result that is close to 0 when the clustering is utterly arbitrary. Accordingly, we also use the ARI \cite{hubert1985comparing}, which corrects the flaws of RI in some way. For both RI and ARI, they take values in the range of 0 to 1, and a value that is close to 1 indicates good agreement between the two partitions. Additionally, we compared clustering results using the VI \cite{meilua2007comparing}. The VI is a criterion derived from information theoretic principles for comparing two clusterings of a data set. The closer its value is to the minimum value of 0, the higher similarity between the true classification situation and the clustering result.

Throughout our simulation studies, we simulate a $100\times100$ data matrix with a checkerboard bicluster structure, where $x_{ij}\stackrel{i.i.d}{\sim}{N(\mu_{rc},\sigma^2)}$ and $\sigma=2$. To assess the performance as the number of column and row clusters varied, we generated data with different numbers of biclusters. The first case is the generation of data using $4\times4$ biclusters, where $\mu_{rc}$ took on one of 21 equally spaced values between -5 and 5, namely $\mu_{rc}{\sim}{Uniform\{-5,-4.5,...,4.5,5\}}$. And the second case is the generation of data using $5\times5$ biclusters, where $\mu_{rc}$ took on one of 25 equally spaced values between -6 and 6, namely $\mu_{rc}{\sim}{Uniform\{-6,-5.5,...,5.5,6\}}$. To investigate the robustness and the effectiveness of these biclustering algorithms, we simulated the heavy-tailed random noise from the following five distribution settings and added it to each element of the data matrix:

\begin{itemize}
    \item Cauchy distribution $C(\gamma,x_0)$ with $\gamma=1.5$ and $x_0=0$
    \item Log-normal distribution ${LN(\mu,\sigma^2)}$ with $\mu=0$ and $\sigma=2$
    \item T distribution ${t(\nu)}$ with degree of freedom $\nu=1$
    \item Pareto distribution ${Pareto(x_{m},\alpha)}$ with scale $x_{m}=1$ and shape $\alpha=2$
    \item Skewed generalized t distribution \cite{theodossiou1998financial} ${sgt(\mu,\sigma^2,\lambda,p,q)}$, where mean $\mu=0$, variance $\sigma^2=q/(q-2)$ with $q=2.5$, shape $p=2$, and skewness $\lambda=0.75$

\end{itemize}

As for the parameters in the weight function, we simply set $k=5$ for both column weight and row weight. Also, we set $\tau=1.345 \hat{\sigma}$, where $\hat{\sigma}$ denotes the median absolute deviation (MAD) estimator of $\mathbf{X}$, in the non-tuning-free method's weight function. We calculated the average mean and standard deviation of RI, ARI, and VI for 50 repeated runs of our proposed methods, COBRA, and spBC. The results for different simulation settings are summarized in Table~\ref{main-table}.

\begin{table}[]
\caption{Results of tuning-free robust convex biclustering (TF-RCBC), non-tuning-free robust convex biclustering (RCBC), COBRA, and spBC in five different distributions of heavy tail noise and two different settings of checkerboard patterns. Each simulation is replicated 50 times, and the means are shown along with the standard deviations in the parenthesis. In each distribution of each setting, the best performance is bolded.}
\renewcommand\arraystretch{1.1}
\setlength{\tabcolsep}{1.0mm}{
\begin{tabular}{ccccccccc}
\hline
                                                    &                                   &            & \multicolumn{2}{c}{\textbf{RI}} & \multicolumn{2}{c}{\textbf{ARI}} & \multicolumn{2}{c}{\textbf{VI}} \\ \hline
\multirow{20}{*}{$\textbf{4} \times \textbf{4}$} & \multirow{5}{*}{\textbf{TF-RCBC}} & Cauchy     & 0.9976              & (0.0026)    & 0.9784              & (0.0239)     & 0.0341              & (0.0330)    \\
                                                    &                                   & log-normal & \textbf{0.9995}     & (0.0007)    & \textbf{0.9955}     & (0.0059)     & \textbf{0.0094}     & (0.0121)    \\
                                                    &                                   & t          & \textbf{0.9995}     & (0.0005)    & \textbf{0.9961}     & (0.0046)     & \textbf{0.0080}     & (0.0088)    \\
                                                    &                                   & Pareto     & \textbf{0.9998}     & (0.0003)    & 0.9979              & (0.0028)     & 0.0048              & (0.0059)    \\
                                                    &                                   & sgt        & \textbf{0.9997}     & (0.0003)    & \textbf{0.9972}     & (0.0030)     & \textbf{0.0059}     & (0.0060)    \\ \cline{3-9} 
                                                    & \multirow{5}{*}{\textbf{RCBC}}    & Cauchy     & \textbf{0.9982}     & (0.0025)    & \textbf{0.9836}     & (0.0230)     & \textbf{0.0284}     & (0.0357)    \\
                                                    &                                   & log-normal & 0.9981              & (0.0029)    & 0.9826              & (0.0266)     & 0.0291              & (0.0395)    \\
                                                    &                                   & t          & 0.9989              & (0.0018)    & 0.9902              & (0.0163)     & 0.0173              & (0.0255)    \\
                                                    &                                   & Pareto     & \textbf{0.9998}     & (0.0002)    & \textbf{0.9980}     & (0.0021)     & \textbf{0.0046}     & (0.0045)    \\
                                                    &                                   & sgt        & 0.9994              & (0.0008)    & 0.9947              & (0.0070)     & 0.0108              & (0.0130)    \\ \cline{3-9} 
                                                    & \multirow{5}{*}{\textbf{COBRA}}   & Cauchy     & 0.9374              & (0.0002)    & 0.0015              & (0.0009)     & 0.7156              & (0.0084)    \\
                                                    &                                   & log-normal & 0.9375              & (0.0001)    & 0.0009              & (0.0006)     & 0.7109              & (0.0069)    \\
                                                    &                                   & t          & 0.9360              & (0.0016)    & 0.0052              & (0.0025)     & 0.7278              & (0.0130)    \\
                                                    &                                   & Pareto     & 0.9522              & (0.0280)    & 0.6074              & (0.1195)     & 0.3487              & (0.0820)    \\
                                                    &                                   & sgt        & 0.9593              & (0.0438)    & 0.7696              & (0.1942)     & 0.1792              & (0.1353)    \\ \cline{3-9} 
                                                    & \multirow{5}{*}{\textbf{spBC}}    & Cauchy     & 0.0624              & (0.0000)    & 0.0000              & (0.0000)     & 1.0000              & (0.0000)    \\
                                                    &                                   & log-normal & 0.0624              & (0.0000)    & 0.0000              & (0.0000)     & 1.0000              & (0.0000)    \\
                                                    &                                   & t          & 0.0624              & (0.0000)    & 0.0000              & (0.0000)     & 1.0000              & (0.0000)    \\
                                                    &                                   & Pareto     & 0.7386              & (0.3202)    & 0.6787              & (0.1920)     & 0.4350              & (0.3195)    \\
                                                    &                                   & sgt        & 0.9848              & (0.0248)    & 0.9203              & (0.1147)     & 0.0572              & (0.0823)    \\ \hline
\multirow{20}{*}{$\textbf{5} \times \textbf{5}$} & \multirow{5}{*}{\textbf{TF-RCBC}} & Cauchy     & 0.9995              & (0.0008    & 0.9931              & (0.0105)     & 0.0109              & (0.0158)    \\
                                                    &                                   & log-normal & \textbf{0.9990}     & (0.0016)    & \textbf{0.9867}     & (0.0221)     & \textbf{0.0196}     & (0.0284)    \\
                                                    &                                   & t          & \textbf{0.9996}     & (0.0009)    & \textbf{0.9942}     & (0.0118)     & \textbf{0.0091}     & (0.0161)    \\
                                                    &                                   & Pareto     & \textbf{0.9997}     & (0.0002)    & 0.9963              & (0.0030)     & 0.0068              & (0.0055)    \\
                                                    &                                   & sgt        & \textbf{0.9998}     & (0.0004)    & \textbf{0.9973}     & (0.0047)     & \textbf{0.0048}     & (0.0081)    \\ \cline{3-9} 
                                                    & \multirow{5}{*}{\textbf{RCBC}}    & Cauchy     & \textbf{0.9996}     & (0.0005)    & \textbf{0.9947}     & (0.0064)     & \textbf{0.0087}     & (0.0093)    \\
                                                    &                                   & log-normal & 0.9983              & (0.0022)    & 0.9769              & (0.0316)     & 0.0330              & (0.0412)    \\
                                                    &                                   & t          & 0.9995              & (0.0014)    & 0.9935              & (0.0197)     & 0.0094              & (0.0257)    \\
                                                    &                                   & Pareto     & \textbf{0.9997}     & (0.0002)    & \textbf{0.9967}     & (0.0027)     & \textbf{0.0058}     & (0.0046)    \\
                                                    &                                   & sgt        & 0.9995              & (0.0007)    & 0.9934              & (0.0112)     & 0.0096              & (0.0137)    \\ \cline{3-9} 
                                                    & \multirow{5}{*}{\textbf{COBRA}}   & Cauchy     & 0.9600              & (0.0001)    & 0.0021              & (0.0012)     & 0.6642              & (0.0083)    \\
                                                    &                                   & log-normal & 0.9600              & (0.0001)    & 0.0012              & (0.0007)     & 0.6616              & (0.0067)    \\
                                                    &                                   & t          & 0.9589              & (0.0015)    & 0.0070              & (0.0034)     & 0.6765              & (0.0189)    \\
                                                    &                                   & Pareto     & 0.9815              & (0.0080)    & 0.7199              & (0.0967)     & 0.2422              & (0.0616)    \\
                                                    &                                   & sgt        & 0.9961              & (0.0038)    & 0.9560              & (0.0429)     & 0.0191              & (0.0180)    \\ \cline{3-9} 
                                                    & \multirow{5}{*}{\textbf{spBC}}    & Cauchy     & 0.0399              & (0.0000)    & 0.0000              & (0.0000)     & 1.0000              & (0.0000)    \\
                                                    &                                   & log-normal & 0.0399              & (0.0000)    & 0.0000              & (0.0000)     & 1.0000              & (0.0000)    \\
                                                    &                                   & t          & 0.0399              & (0.0000)    & 0.0000              & (0.0000)     & 1.0000              & (0.0000)    \\
                                                    &                                   & Pareto     & 0.8290              & (0.2748)    & 0.7235              & (0.1481)     & 0.3388              & (0.2904)    \\
                                                    &                                   & sgt        & 0.9847              & (0.0162)    & 0.8531              & (0.1409)     & 0.0899              & (0.0787)    \\ \hline
\end{tabular}}
\label{main-table}
\end{table}

Table~\ref{main-table} demonstrates the average mean and standard deviation of RI, ARI, and VI in 50 replicated runs. To make things clear, although all of our simulation settings are heavy-tailed distributions, the level of heaviness is different. Accordingly, we can separate all five heavy-tailed distribution settings into two groups: low noise group and high noise group. Low noise group contains Pareto distribution and skewed generalized t distribution, while high noise group contains Cauchy distribution, log-normal distribution, and t distribution. In low noise scenarios, even COBRA and spBC, which don't have robust properties, can have a not bad performance. However, when the tails of the noise are becoming increasingly heavier, like in high noise group, COBRA and spBC will hardly be able to recover checkerboard patterns and consequently collapse. Despite the poor performance of COBRA and spBC, our proposed RCBC shows a consistently desirable and robust performance in both low and high noise scenarios. Also, we can see that the effectiveness of tuning-free and non-tuning-free RCBC is very close and similar to each other. Later, we will show an image example of different types of the collapse of COBRA and spBC, along with our proposed RCBC's good performance.

\begin{figure}[htbp]
\centering
\subfigure
{
    \begin{minipage}[b]{.22\linewidth}
        \centering
        \includegraphics[width=\textwidth]{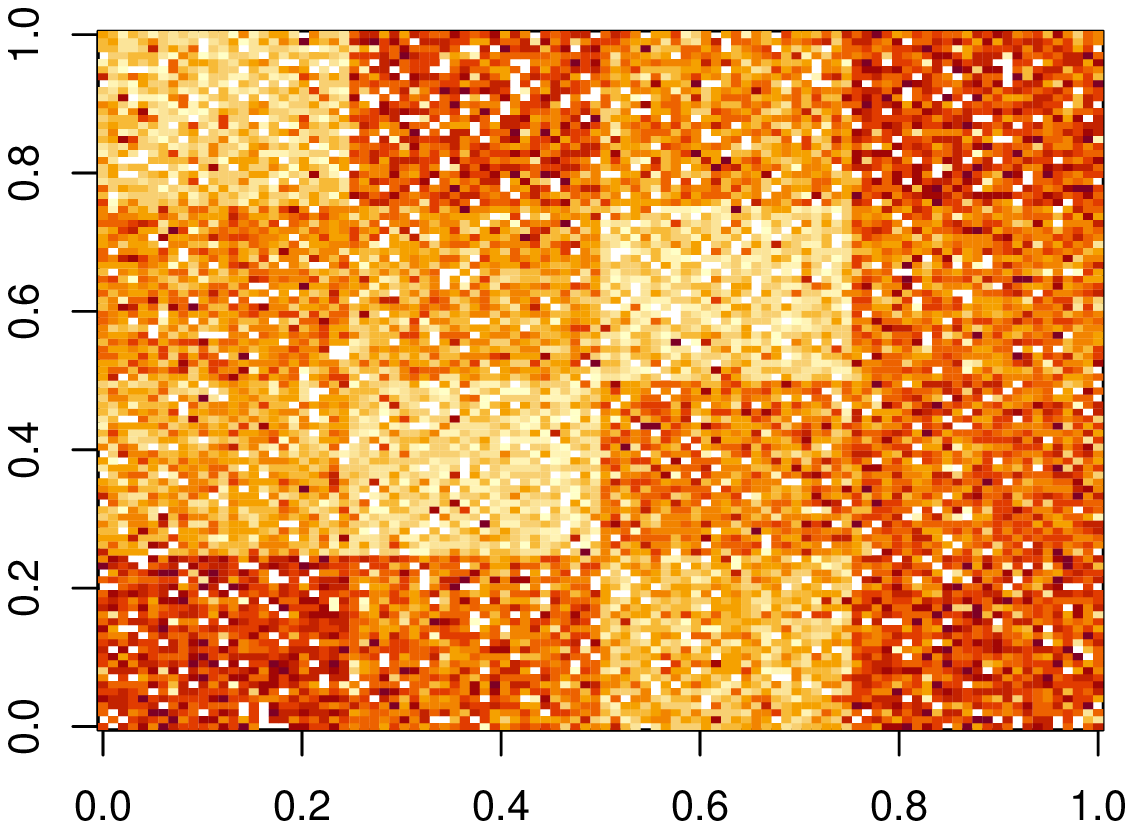}
        \centerline{(a)}
        \label{fig-sim-a}
    \end{minipage}
}
\subfigure
{
 	\begin{minipage}[b]{.22\linewidth}
        \centering
        \includegraphics[width=\textwidth]{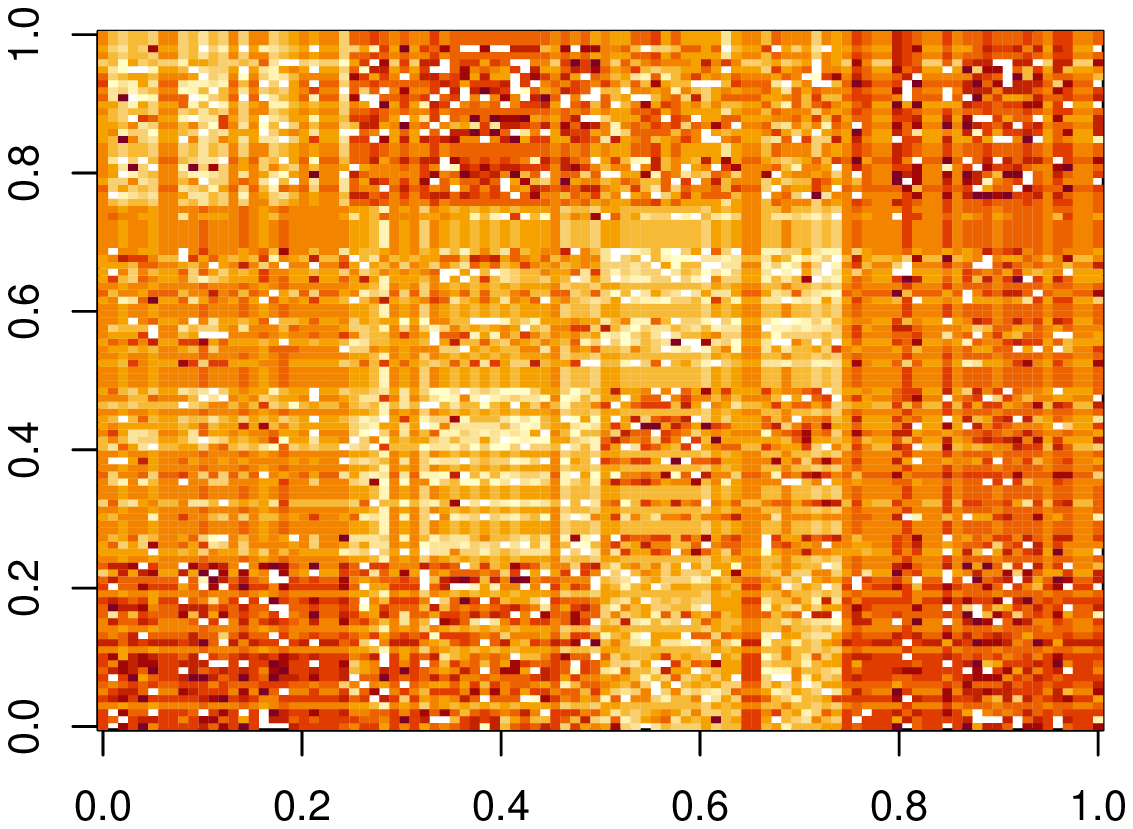}
        \centerline{(b)}
        \label{fig-sim-b}
    \end{minipage}
}
\subfigure
{
 	\begin{minipage}[b]{.22\linewidth}
        \centering
        \includegraphics[width=\textwidth]{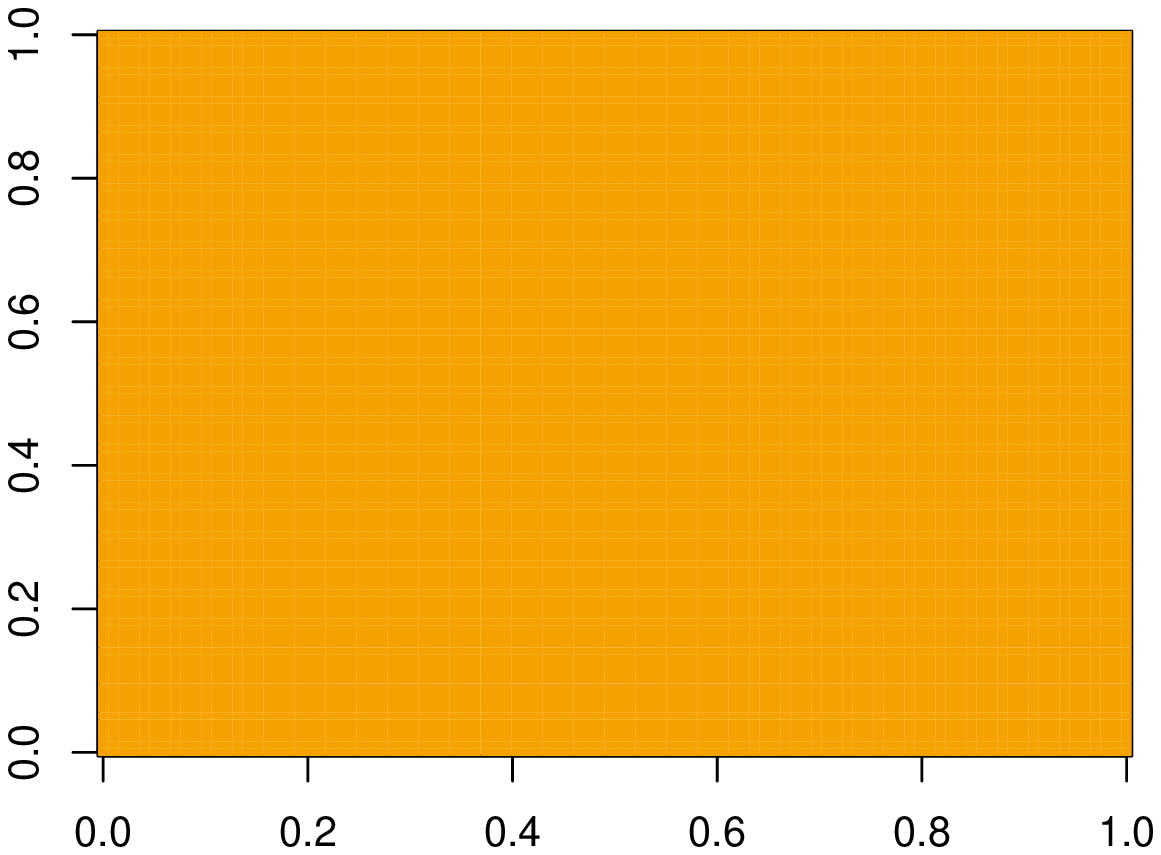}
        \centerline{(c)}
        \label{fig-sim-c}
    \end{minipage}
}
\subfigure
{
 	\begin{minipage}[b]{.22\linewidth}
        \centering
        \includegraphics[width=\textwidth]{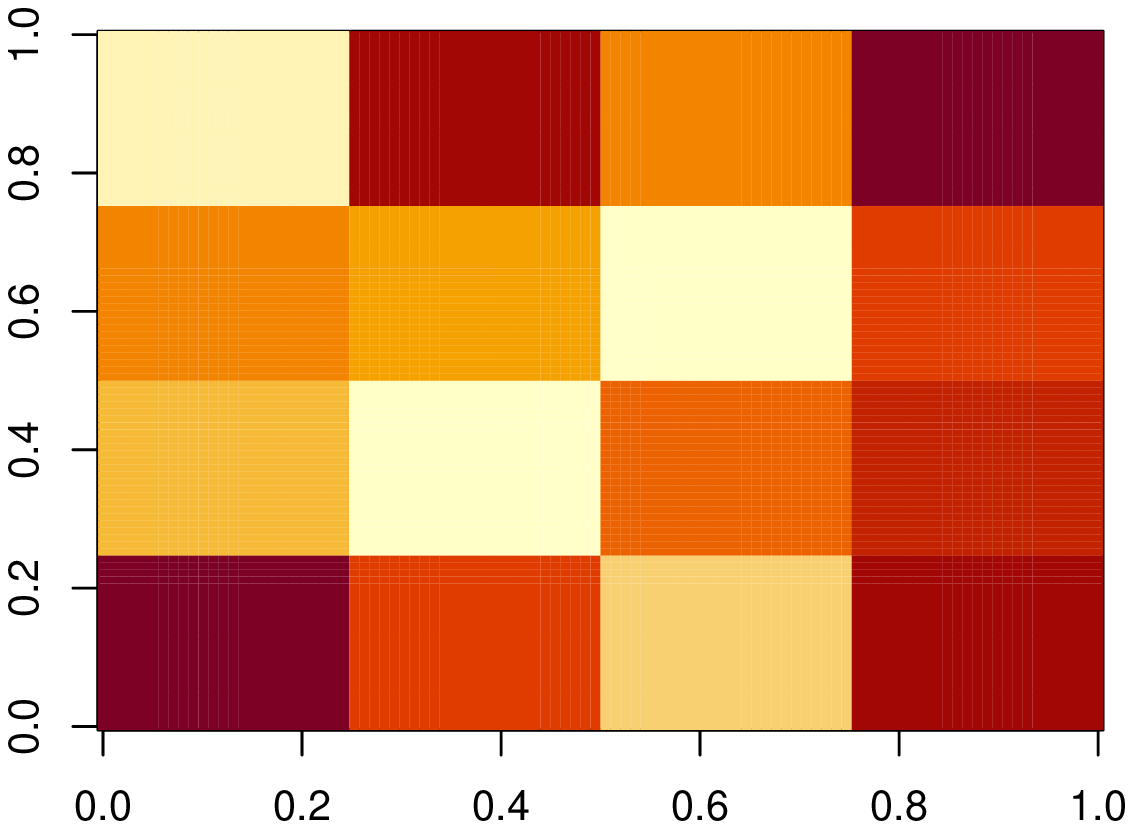}
        \centerline{(d)}
        \label{fig-sim-d}
    \end{minipage}
}
\caption{Example of the performance of four compared algorithms when the data matrix is contaminated by t distribution with degree of freedom being 1. (a) Original data matrix. (b) Result matrix of COBRA. (c) Result matrix of spBC. (d) Result matrix of RCBC}
\label{fig_sim}
\end{figure}

Figure~\ref{fig_sim} can give us an intuitive illustration of COBRA and spBC's bad performance when facing the data matrix contaminated by extremely heavy-tailed distributions. The estimated mean matrix of COBRA (Figure~\ref{fig-sim-b}) is almost the same as the original data matrix (Figure~\ref{fig-sim-a}), which represents the collapse of COBRA in high noise scenario. Moreover, the estimated mean matrix of spBC (Figure~\ref{fig-sim-c}) is literally a constant matrix, which represents another type of collapse of spBC. These two types of collapse of the biclustering algorithms can also be reflected by Table~\ref{main-table} where the ARI of COBRA is close to 0 and the ARI of spBC is utterly 0 in high noise scenarios. Additionally, Figure~\ref{fig_sim} also demonstrates how our proposed RCBC perfectly recovers the underlying checkerboard patterns in this data matrix.

\begin{figure}
    \centering
    \includegraphics[width=.9 \textwidth]{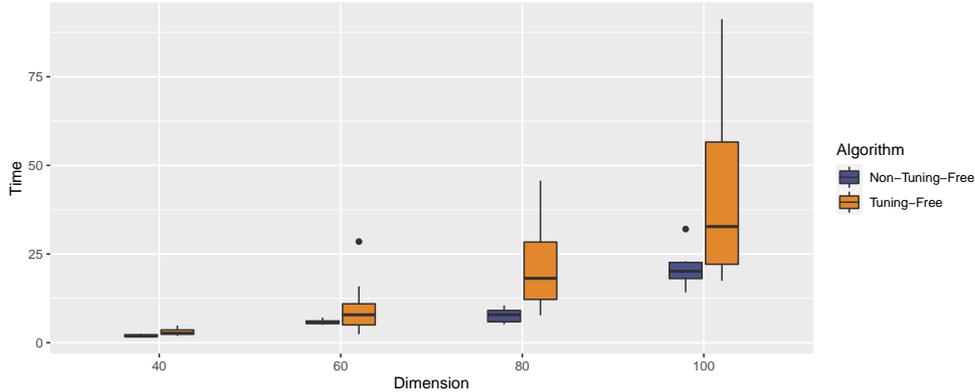}
    \caption{Boxplot of the speed of our proposed tuning-free method and non-tuning-free method. Row label represents the time needed for running an algorithm in seconds. Column label represents the number of rows ($n$) and columns ($p)$ of the data matrix at the same time. Note that in this simulation, all the data matrices are square matrices, i.e. the number of rows equals the number of columns.}
    \label{fig-boxplot}
\end{figure}

We also compare the speed of our proposed tuning-free method and non-tuning-free method. Figure~\ref{fig-boxplot} is a boxplot reflecting the speed of two algorithms which are rerun 10 times with $LN(0,2^2)$ noise. Although the tuning-free method is a little bit slower than the non-tuning-free one because of some extra iterative procedures in the algorithm, it does save a lot of time when selecting the optimal parameters by CV (cross-validation). The CV procedure with missing data is very computationally demanding. Therefore, reducing the number of tuning parameters from two to one is great computational simplicity and efficiency.

\section{Application}
\label{sec-app}

We now consider applying our proposed RCBC to real-life microarray data. In this application domain, we are trying to simultaneously cluster genes and cancer subtypes. To be specific, biclustering algorithms help to identify genes that are significantly expressed for certain cancer subtypes. Lung cancer gene expression data is a very popular and classic example in biclustering problems \cite{lee2010biclustering,tan2014sparse,chi2017convex}. The dataset contains 56 samples and 12625 genes. All 56 subjects can be divided into the following four subgroups:

\begin{itemize}
    \item Normal: normal subjects(Contains 17 samples)
    \item Carcinoid: pulmonary carcinoid tumours (Contains 20 samples)
    \item Colon: colon metastases (Contains 13 samples)
    \item Small Cell: small cell carcinoma (Contains 6 samples)
\end{itemize}

\begin{figure}
    \centering
    \includegraphics[width=.75 \textwidth]{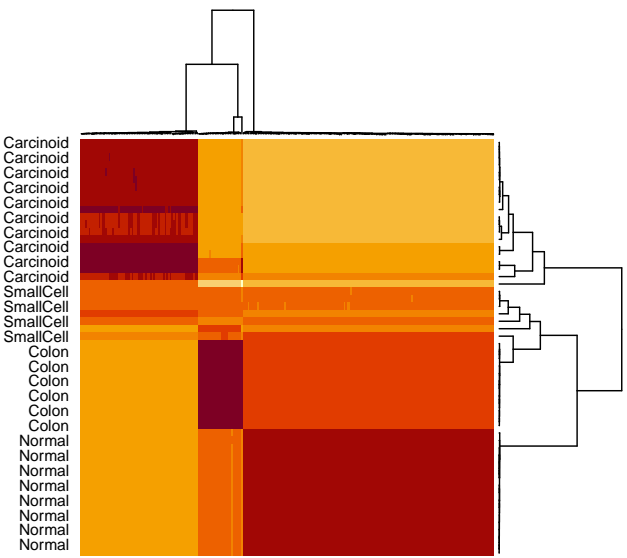}
    \caption{Heatmap of the result from our proposed RCBC on a subset of the lung cancer data consisting of the 250 genes with the highest variance. Column label is the cancer subtype.}
    \label{fig-gene}
\end{figure}

We have selected the 250 genes with the greatest variance. When it comes to performing our algorithm on the data, we set $k=10$ for row weight and $k=8$ for column weight. A heatmap of the result of our proposed method is shown in Figure~\ref{fig-gene}.

Figure~\ref{fig-gene} vividly illustrates our success in finding the local correlation between cancer subtypes and genes. Checkerboard patterns emerge, we can easily identify three biclusters and assign each of them to normal subjects, pulmonary carcinoid tumours, and colon metastases respectively. To elaborate, the genes in the lower part have a larger mean in pulmonary carcinoid tumours, which may be interpreted as highly expressed in this cancer subtype. Likewise, the genes in the middle and upper part have larger means in colon metastases and normal subjects respectively, which may represent the high expression in these two subtypes respectively.

\begin{figure}[htbp]
\centering
\subfigure
{
    \begin{minipage}[b]{.45\linewidth}
        \centering
        \includegraphics[width=\textwidth]{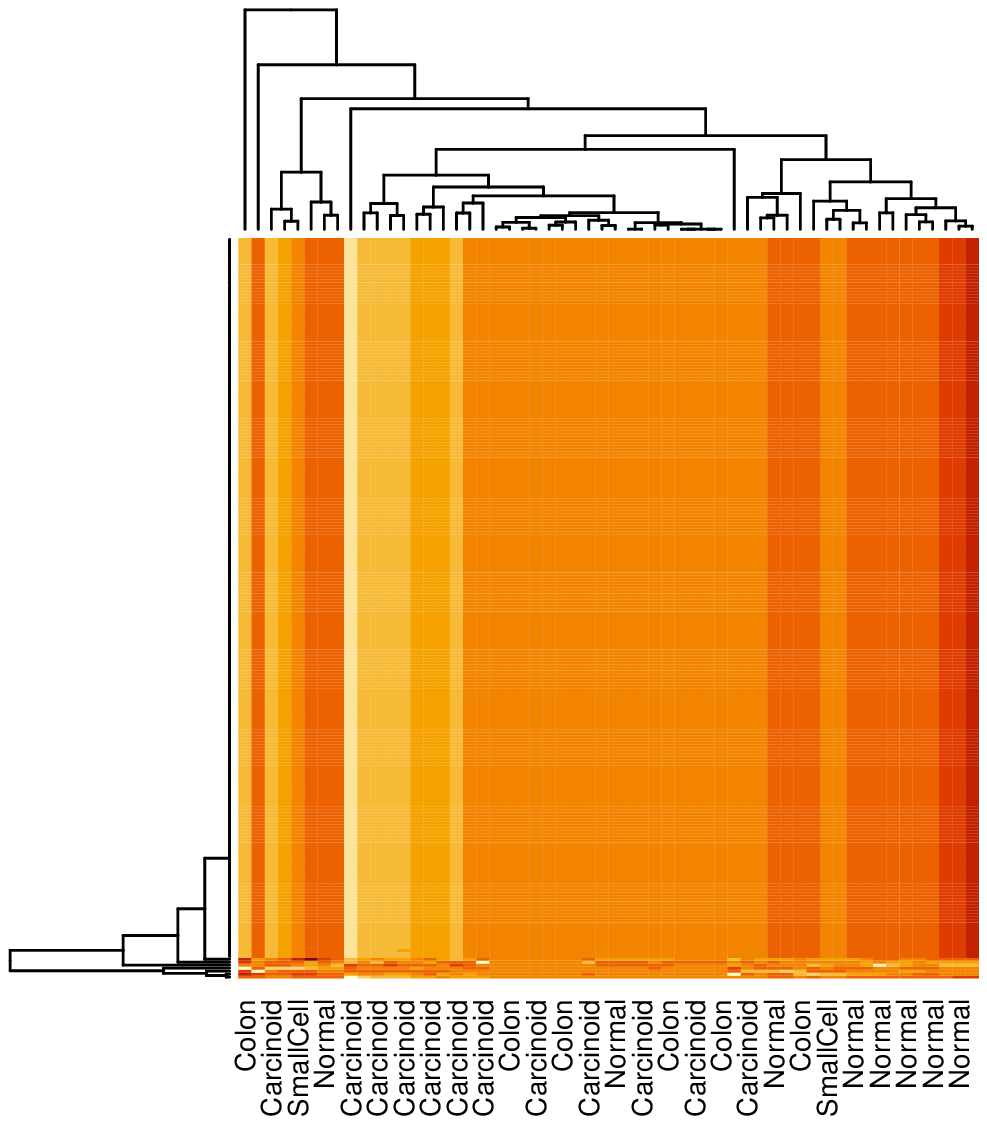}
        \centerline{(a)}
    \end{minipage}
}
\subfigure
{
 	\begin{minipage}[b]{.45\linewidth}
        \centering
        \includegraphics[width=\textwidth]{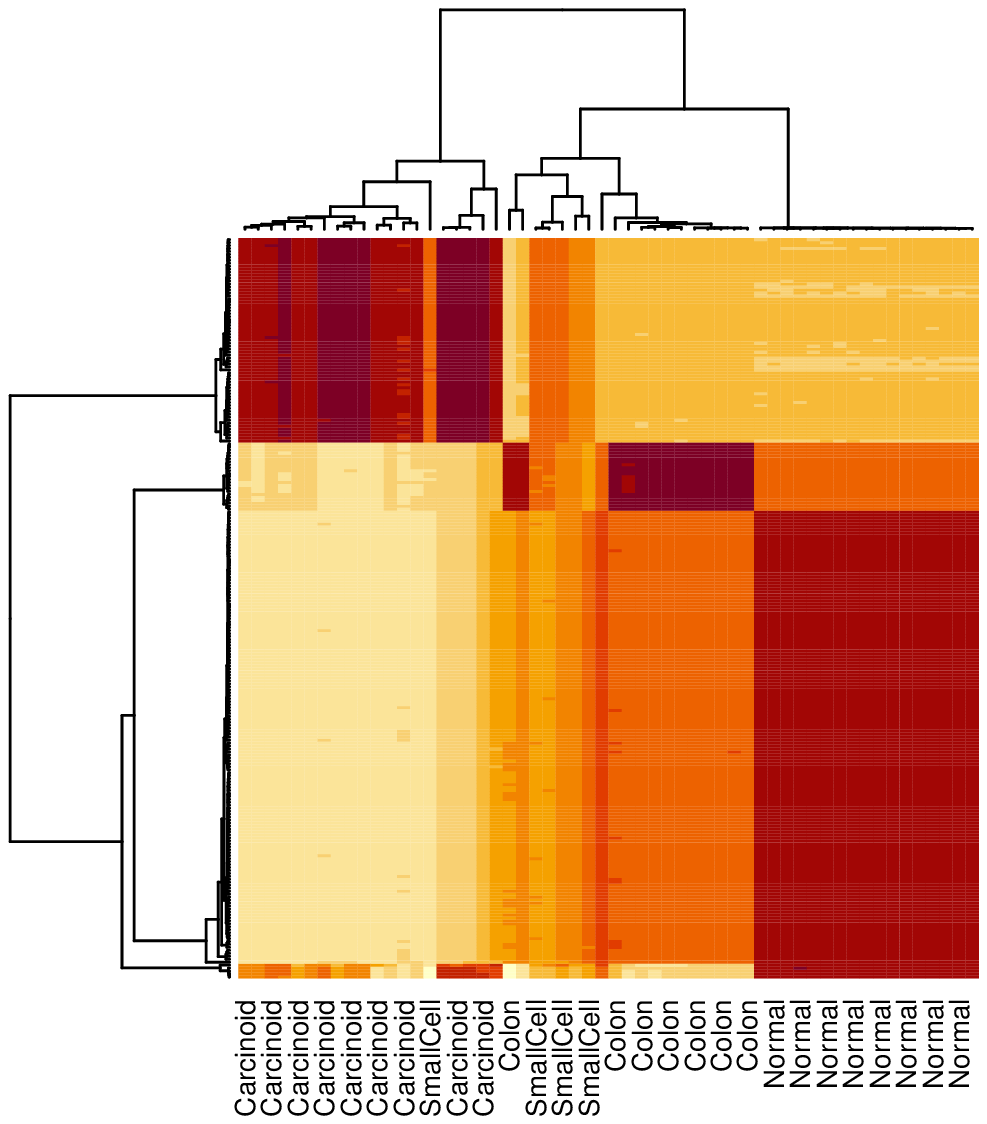}
        \centerline{(b)}
    \end{minipage}
}

\caption{Comparison of the performance of COBRA and RCBC on lung cancer gene expression data with heavy-tailed noise. Column label is the cancer subtype. (a)Heatmap of COBRA. (b)Heatmap of RCBC. }
\label{fig-gene-t1}
\end{figure}

Moreover, in order to verify the robust feature of RCBC in the application, we add $t(1)$ noise to the lung cancer gene expression data. Then, we implement COBRA and RCBC both with $k=5$ for row weight and $k=8$ for column weight on this same contaminated data. Hopefully, we will still see our proposed RCBC more or less outperform COBRA. We report the result of these two algorithms in Figure ~\ref{fig-gene-t1}.

From Figure~\ref{fig-gene-t1}, we can see that while COBRA almost collapse on heavy-tailed gene expression data, RCBC could still have a desirable performance on the same dataset. We can still easily identify biclusters that demonstrate the local correlation of lung cancer subtype and gene expression. Furthermore, the detailed correlation between genes and cancer subtypes is almost the same as the result on the original uncontaminated dataset. Therefore, we can say that although the data contains heavy-tailed noise, RCBC is still robust and accurate.

\section{Discussion}
\label{sec-dis}

In this paper, we proposed a robust version of convex biclustering (RCBC) to deal with the heavy tail noise in real life. The robust property is achieved by substituting the square loss with the Huber loss. Additionally, we proposed a tuning-free method for selecting the robustification parameter $\tau$ in order to speed up the procedure of finding the optimal parameters. We've shown that in extremely heavy tail scenarios, our proposed algorithm outperforms its ancestor COBRA to a large extent. Also, we manifested the feasibility of applying our algorithm to real-life biomedical data.

Despite the excellent result of our simulation study, there are some aspects to be improved. First, the speed of our algorithm is not very competitive compared to other biclustering algorithms. Accordingly, there could be some improvement in the derivation of optimization. Second, our proposed method does not allow for overlapping biclusters, i.e., each element in the data matrix can only be in one bicluster. However, although some advantages exist for overlapping biclusters in certain contexts \cite{madeira2004biclustering}, it is very complex and hard to interpret \cite{tan2014sparse}.

\section*{Disclosure statement}

No potential conflict of interest was reported by the authors.

\section*{Funding}

This work was supported by the Technology Research Project of Education Department of Jiangxi Province under Grant No. GJJ210535; National Natural Science Foundation of China
under Grant No. 12161042; and China Postdoctoral Science Foundation under Grant No. 2019M662262.

\bibliographystyle{tfnlm.bst}
\bibliography{interactnlmsample}

\appendix
\section{Derivation of (\ref{formula-one})}
\label{appendix}

In (\ref{formula-lag}), we already have the augmented Lagrangian function. We recast the function here again:

$$\begin{aligned}
        F(\mathbf{U},\mathbf{W},\mathbf{V},\mathbf{Y},\mathbf{Z})&=\mathbf{L}_\tau(\mathbf{X}-\mathbf{W})+\lambda\sum_{i<j}w_{ij}||\mathbf{V}_{ij}||_2\\
        &+\frac{\rho}{2}||\mathbf{V-EU+Y}||_F^2+\frac{\rho}{2}||\mathbf{W-U+Z}||_F^2
    \end{aligned}
$$

By the ADMM algorithm, the update for $\mathbf{U},\mathbf{W},\mathbf{V},\mathbf{Y}$ and $\mathbf{Z}$ can be derived as follows: \\

\leftline{\textbf{a) Derivation for the update of} $\mathbf{U}$:}

$$\mathbf{U}^{(m)}=\mathop{\arg\min}_{\mathbf{U}}F\left(\mathbf{U},\mathbf{W}^{(m-1)},\mathbf{V}^{(m-1)},\mathbf{Y}^{(m-1)},\mathbf{Z}^{(m-1)}\right)$$ This optimization problem can be write as:

$$\mathop{\min}_{\mathbf{U}} \frac{\rho}{2}||\mathbf{V-EU+Y}||_F^2+\frac{\rho}{2}||\mathbf{W-U+Z}||_F^2$$ The solution to this problem is trivial by taking the matrix derivative of $\mathbf{U}$:

$$\mathbf{U}=\left(\mathbf{E}^T\mathbf{E}+\mathbf{I}\right)^{-1}\left[\mathbf{E}^T(\mathbf{V}+\mathbf{Y})+\mathbf{W}+\mathbf{Z}\right]$$

\leftline{\textbf{b) Derivation for the update of} $\mathbf{W}$:}

$$\mathbf{W}^{(m)}=\mathop{\arg\min}_{\mathbf{W}}F\left(\mathbf{U}^{(m)},\mathbf{W},\mathbf{V}^{(m-1)},\mathbf{Y}^{(m-1)},\mathbf{Z}^{(m-1)}\right)$$ Ignore the uncorrelated variables, and the problem is equivalent to 

$$\mathop{\min}_{\mathbf{W}} \mathbf{L}_\tau(\mathbf{X}-\mathbf{W})+\frac{\rho}{2}||\mathbf{W-U+Z}||_F^2$$ We can solve the above problem element-wise:

$$\mathop{\min}_{W_{ij}} \mathcal{L}_\tau(X_{ij}-W_{ij})+\frac{\rho}{2}(W_{ij}-U_{ij}+Z_{ij})^2$$ (i) When $|X_{ij}-W_{ij}| \leq \tau$:

\leftline{The Huber loss will become square error loss, so the corresponding problem will be:}

$$\mathop{\min}_{W_{ij}} \frac{1}{2}(X_{ij}-W_{ij})^2+\frac{\rho}{2}(W_{ij}-U_{ij}+Z_{ij})^2$$ Take the derivative of $W_{ij}$ and set the first order derivative of the above formula to be 0 can we get

$$W_{ij}=\frac{X_{ij}+\rho(U_{ij}-Z_{ij})}{1+\rho}$$ Take this result into the condition $|X_{ij}-W_{ij}| \leq \tau$, we can have a more useful condition:

$$\frac{\rho}{1+\rho}\left|X_{ij}-\left(U_{ij}^{(m)}-Z_{ij}^{(m-1)}\right)\right|\leq \tau$$ (ii) When $|X_{ij}-W_{ij}|>\tau$, or equivalent to $\frac{\rho}{1+\rho}\left|X_{ij}-\left(U_{ij}^{(m)}-Z_{ij}^{(m-1)}\right)\right|> \tau$:

\leftline{The corresponding optimization problem will become:}

$$\mathop{\min}_{W_{ij}} \tau|X_{ij}-W_{ij}|+\frac{\rho}{2}(W_{ij}-U_{ij}+Z_{ij})^2$$ We can rewrite it into the form that can be easily solved by the soft-thresholding operator:

$$\mathop{\min}_{W_{ij}-X_{ij}} \frac{1}{2}\left[(W_{ij}-X_{ij})-(U_{ij}-Z_{ij}-X_{ij})\right]^2+\frac{\tau}{\rho}|W_{ij}-X_{ij}|$$ Thus 

$$W_{ij}=X_{ij}+soft\left(U_{ij}-Z_{ij}-X_{ij},\frac{\tau}{\rho}\right)$$ where $soft(a,b)=sign(a)\mathcal{\max}(|a|-b,0)$ is the the soft-thresholding operator. \\

\leftline{\textbf{c) Derivation for the update of} $\mathbf{V}$:}

$$\mathbf{V}^{(m)}=\mathop{\arg\min}_{\mathbf{V}}F\left(\mathbf{U}^{(m)},\mathbf{W}^{(m)},\mathbf{V},\mathbf{Y}^{(m-1)},\mathbf{Z}^{(m-1)}\right)$$ By ignoring the uncorrelated variables and rewrite the problem row-wise, we have:

$$\mathop{\min}_{\mathbf{V}_{ij}} \frac{\lambda w_{ij}}{\rho}||\mathbf{V}_{ij}||_2+\frac{1}{2}||\mathbf{V}_{ij}-(\mathbf{U}_i-\mathbf{U}_j)+\mathbf{Y}_{ij}||_2^2$$ which can be viewed as a group lasso problem with the following solution:

$$\mathbf{V}_{ij}=\left[1-\frac{\lambda w_{ij}}{\rho ||\mathbf{U}_i-\mathbf{U}_j-\mathbf{Y}_{ij}||_2}\right]_{+}\left(\mathbf{U}_i-\mathbf{U}_j-\mathbf{Y}_{ij}\right)$$ where $[a]_+=\mathcal{\max}(a,0)$ \\

\leftline{\textbf{d) Derivation for the update of} $\mathbf{Y}$ \textbf{and} $\mathbf{Z}$:}

$$\begin{aligned}
    \mathbf{Y}_{ij}^{(m)}&=\mathbf{Y}_{ij}^{(m-1)}-\rho \left(\mathbf{U}_i^{(m)}-\mathbf{U}_j^{(m)}-\mathbf{V}_{ij}^{(m)}\right) \\
    \mathbf{Z}^{(m)}&=\mathbf{Z}^{(m-1)}-\rho \left(\mathbf{U}^{(m)}-\mathbf{W}^{(m)}\right)
\end{aligned}
$$

\end{document}